# Changes in Fluctuation Waves in Coherent Airflow Structures with Input Perturbation

OSAMA A. MARZOUK
Department of Engineering Science and Mechanics
Virginia Polytechnic Institute and State University
219 Norris Hall (Mail Code 0219)
Blacksburg, VA 24061, USA
omarzouk@vt.edu

*Abstract:* - We predict the development and propagation of the fluctuations in a perturbed ideally-expanded air jet. A non-propagating harmonic perturbation in the density, axial velocity, and pressure is introduced at the inflow with different frequencies to produce coherent structures in the airflow, which are synchronized with the applied frequency. Then, the fluctuations and acoustic fields are predicted for each frequency by integrating the axisymmetric linearized Euler equations. We investigate the effect of the perturbation frequency on different characteristics of the pressure and velocity waves. The perturbation frequency can be used to alter the propagation pattern and intensity of the waves and mitigate the noise levels at certain zones.

*Key-Words:* - Perturbation, Fluctuation, Waves, Acoustic, Coherent, Air, Jet, Synchronized

## 1 Introduction

The word 'sound' typically refers to audible pressure fluctuations in the ambient air. Audible acoustic waves can be generated by the vibrations of solid surfaces such as the vibrating strings of violins and harps, and the vibrating diaphragms of loudspeakers. They can also be generated as a result of unsteady addition or removal of heat as in singing flames [1]. Spherically-expanding flames act as a monopole source of sound [2]. We here consider pressure waves due to turbulent mixing in air jet, which is illustrated in Fig.1. The development of high-speed transport aircrafts is retarded by the community noise standards set by the Federal Aviation Regulations (FAR) and Joint Aviation Authorities (JAA). The high-speed jet flows emanated from turbojet engines are the main source of noise of commercial aircrafts. Compliance with the FAR and JAR noise standards is remarkably affecting the progress in high-speed air transportation. Jet engines with bypass (cold jet passes through the fan in addition to core hot stream that from the jet engine) have less noise than the conventional engines, but at the expense of increased drag and weight. Passive noise control devices for these jets include deeply-corrugated jet exit and multinozzle noise suppressor, but they are not suitable for high-bypass-ratio engines. Applying perturbations of small amplitude near the jet exit has minimal effects on the thrust but can alter the propagation of acoustic waves. This may lead to an active control technique but research is needed to examine this idea. We conduct a preliminary study on the effect of the varying the frequency of such perturbation on the intensity and pattern of the pressure field associated with an axisymmetric air jet.

Attempts toward noise estimation from air jets go back more than fifty years. Some early measurements of the intensity of far field noise from turbulent air jets were performed by Morely [3]. Analytical, experimental, and numerical approaches have been used to assess flow noise. Analytical approaches include Lighthill's acoustic analogy [4], which enables the prediction of jet noise if the source of the sound is known a priori. However, detailed experimental measurements of the source and its retarded time would be very difficult and expensive. Numerical approaches include the linear stability theory, direct numerical simulation, and large eddy simulation. Linear stability theory considers only instability modes, which limits the range of problems that can be solved accurately. Direct numerical simulation [5, 6] solves, without simplifying modeling, the time-dependent, compressible Navier-Stokes equations that govern both sound generation and propagation. This method is very expensive computationally. In addition, there is a possibility that the order of magnitude of the acoustic field be of comparable order of the numerical errors of the resolved (total) field. There are few reported studies where this approach was used to predict the noise, such as the study conducted [7, 8]. The large eddy simulation is another numerical approach with less computational demands. In this approach, the flow field is decomposed into large-scales and small-scales (also called subgrid scales), both are time-dependent. The former are captured directly, whereas the latter are modeled. The noise generated by the small-scale components is not captured directly. Also, the acoustic field can be blurred by the numerical errors implied in calculating the flow variables if the magnitude of the acoustic field is very small compared to the total field. Bodony and Lele [9] conducted a review of the status of jet noise predictions using large eddy simulation, considering both subsonic and supersonic jets. Recent hybrid techniques [10, 11], utilize two different numerical approaches. The key point is to split the problem and apply each approach to the suitable portion so that the overall technique becomes advantageous, in terms of time or accuracy, to the case of using either approach individually. The linearized Euler equations [12-14] govern the inviscid small-amplitude fluctuations. It is suitable for aeroacoustic problems of free shear layers and jets because the acoustic field (or the pressure fluctuations) are of small amplitude, and viscous effects are not significant in the linear region [15, 16].

In addition to the extensive research on the numerical prediction of noise, there is interest in investigating the behavior of externally-perturbed jets and their acoustic fields. Chan [17] considered a perturbed axisymmetric air jet at Reynolds number $2.6 \times 10^5$. In the experimental setup, air was stored in a pressurized tank and was flowing through a valve to a settling cylindrical chamber. The chamber was connected to a parallel nozzle through a contraction section. The periodic perturbation was implemented through a loudspeaker and horn assembly in the upstream end of the chamber and in line with the nozzle. The perturbation amplitude was 125 dB (Ref 20 µPa) at the center of the nozzle exit when the jet was running. It was observed that the wavenumbers of the acoustic waves increase monotonically with the perturbation frequency. McLaughlin et al. [18] used hot-wires and microphones to measure the flow and acoustic fields for perturbed supersonic air jet with Reynolds numbers varying from 8,000 to 107,000. The jet was excited by a glow discharge located at a single point tungsten electrode, approximately 2mm from the jet exit. The periodic glow was generated by applying AC voltage of peak-to-peak 800 V with −450 V DC bias. The perturbation propagated with a higher speed than the ambient speed of sound and grew like instability waves. The study showed that large-scale instability is the dominant source of noise. Troutt [19] used a similar approach to perturb air jet from an axisymmetric supersonic nozzle in order to generate and study coherent fluctuations by strengthening the harmonic component in the flow. According to Morris and Tam [20], such periodic perturbations can be thought of as induced by flow turbulence as in the case of unperturbed jets. A full solution of the jet problem should include the internal flow in the nozzle and the engine itself, in case of hot jet. This is because the exact nature of the perturbation in the jet core is governed by these conditions. However, this makes the problem very complicated and impractical and requires tremendous computational resources [21]. Therefore, approximations using perturbation signals are useful in this case to mimic realistic situations. Lew et al. [22] solved the unsteady, Favre-filtered, compressible large eddy simulation equations to study the effect of randomized perturbations in subsonic jet at Reynolds number 100,000. Although they used a program code with capability of utilizing classical and localized dynamic Smagorinsky subgrid-scale model, they did not use any subgrid scale modeling in order to shorten the runtimes. The inflow perturbations were applied as induced velocities from a vortex ring. The focus was on the sensitivities of the flow development and jet noise to the number of modes. They found that eliminating the first few modes causes longer potential core and an increase of 1-2 dB in the sound pressure levels in the far field.

We consider a supersonic axisymmetric air jet described by a set of analytical and conservation equations, and numerically apply harmonic pressure, density, and velocity perturbations in the inflow (near the virtual nozzle location) at different frequencies so that the fluctuations are synchronized with the applied frequency as the parameter of the perturbation. We solve the linearized Euler equations using the finite-difference method and non-reflecting boundary conditions to predict and analyze the propagation of these synchronized fluctuations and the associated pressure and velocity waves, focusing of the effect of the perturbation frequency on these waves.

## 2 Jet Properties

The flow is decomposed into two fields. The first one is time-averaged (mean), whereas the second is time-dependent (fluctuations). The mean flow is provided by an analytical approach proposed by Tam and Burton [23], applied to measurements of supersonic ideally-expanded air jet by Trout and McLaughlin [24]. The fluctuations are resolved by solving the linearized Euler equations. One of the favorable features in this splitting is that the computation of the fluctuations is not very sensitive to the mean field [25]. In addition, the phases of computing the fluctuations field and mean field are independent, thereby giving a big amount of flexibility in the overall process. The jet properties at the nozzle exit are given in Table 1. We should mention that the low density is mandatory to decrease the dynamic pressures to levels that are suitable, with the high speeds of the airflow, for the probes of the hot-wire anemometer that were used in the measurements. The subminiature probes had wires diameter of 5 μm and were positioned perpendicular to the axial direction.

**Table 1. Properties of the air jet at the nozzle exit**

| | |
|---|---|
| Diameter | 10 mm |
| Pressure | 5,060 N/m$^2$ |
| Total pressure | 46,270 N/m$^2$ |
| Temperature | 156.2 °K |
| Total temperature | 294°K |
| Density | 0.1114 Kg/m$^3$ |
| Velocity | 525 m/s |
| Speed of sound | 250 m/s |
| Reynolds number | 70,000 |

A Gaussian radial distribution is used for the axial velocity $U$ outside the core. The measured axial profile of the mixing layer size is related to the peak $U$ of the profile through the conservation of axial momentum flux. Invoking boundary-layer type approximation to the axisymmetric time-averaged Euler equations shows that mean flow pressure can be taken uniform in the jet. Under such assumption, the continuity equation for the axisymmetric flow reduces to that of the incompressible flow, which is

$$\frac{\partial U}{\partial x} + \frac{1}{r}\frac{\partial (r V)}{\partial r} = 0 \qquad (1)$$

From Eq.(1), the radial velocity $V$ is obtained at any axial distance $x'$ from

$$V(x',r) = -\frac{1}{r}\int_0^r \frac{\partial U}{\partial x} r' dr' \qquad (2)$$

where $r$ is the radial coordinate and $r'$ is a dummy radial variable. The above expressions for $U$ and $V$ are used to describe the mean flow up to a maximum radius $r_{max}(x)$, which is three times the radius of the mixing layer, and its axial profile is given in Fig.2 in terms of the jet radius at the nozzle exit $R_j$. Beyond this value, the following expressions are used

$$U = 0$$
$$V = -\frac{V_\infty(x)}{r} \qquad (3)$$

where $V_\infty(x)$ is the radial velocity at the radial boundary of the domain, which is at 16 jet diameters.

The total temperature $T_o$ is constant, which is used to obtain the profile of the static temperature in terms of the velocities as

$$T = T_o - \frac{U^2 + V^2}{2\,Cp} \qquad (4)$$

where $Cp$ is the specific heat of air at constant pressure, which is 1,004 J/kg-K. The relation in Eq.(4) is an energy-conservation form balancing the kinetic energy and enthalpy. The equation of state $P = \bar{\rho}\,R_{air}\,T$ is used to obtain the density in the mean flow $\bar{\rho}$, where $P$ is the pressure and $R_{air}$=287 J/kg-K is the gas constant of air. The local Mach number is defines as

$$M = \frac{U}{\sqrt{\gamma\,R_{air}\,T}} \qquad (5)$$

where $\gamma$ =1.4 for air. In Fig.3, the radial profiles of $M$ are shown at different values of $x$. In Fig.4, the axial profiles of $V$ relative to the jet velocity at the nozzle exit $U_j$ are shown at different values of $r$. Whereas $V$ is positive at small values of $r$, it is negative at larger values (e.g., larger than $5\,R_j$), reflecting entrainment-flow effects.

## 3 Fluctuations Equations

To linearize the Euler equations about the mean solution, the flow variables are decomposed according to

$$a(x,r,t) = A(x,r) + a'(x,r,t) \qquad (6)$$

where $a$ is a generic variable, such as the density or axial velocity, $A$ is the time-averaged value of $a$, and $a'$ is its fluctuating component.

Upon applying this decomposition to the Euler equations and subtracting the time-averaged equations, we get the linearized version governing the fluctuations. For axisymmetric flow, these equations, in vector from, are

$$\frac{\partial q'}{\partial t} + \frac{\partial F'}{\partial x} + \frac{1}{r}\frac{\partial (rG')}{\partial r} = \frac{1}{r}S' \qquad (7)$$

where $q' = [\rho'\ \tilde{u}\ \tilde{v}\ \tilde{e}]^T$ is the fluctuating conservative variables, $F' = [f_1\ f_2\ f_3\ f_4]^T$ is the fluctuating momentum flux in $x$ direction, $G' = [g_1\ g_2\ g_3\ g_4]^T$ is the fluctuating momentum flux in $r$ direction, and $S' = [0\ 0\ p'\ 0]^T$ is the fluctuating source term that arises in the cylindrical coordinates. The following relations and definitions augment the expression in Eq.(7):

$$\begin{aligned}
\tilde{u} &\equiv (\rho u)' \\
\tilde{v} &\equiv (\rho v)' \\
\tilde{e} &\equiv (\rho e)'
\end{aligned} \qquad (8)$$

$$\begin{aligned}
f_1 &= \tilde{u} \\
f_2 &= -\rho'\,U^2 + 2\,\tilde{u}\,U + p' \\
f_3 &= -\rho'\,U\,V + \tilde{v}\,U + \tilde{u}\,V \\
f_4 &= U\,(p' + \tilde{e}) + (\tilde{u} - \rho'\,U)(P/\bar{\rho} + E)
\end{aligned} \qquad (9)$$

$$\begin{aligned}
g_1 &= \tilde{v} \\
g_2 &= -\rho'\,U\,V + \tilde{v}\,U + \tilde{u}\,V \\
g_3 &= -\rho'\,V^2 + 2\,\tilde{v}\,V + p' \\
g_4 &= V(p' + \tilde{e}) + (\tilde{v} - \rho'\,V)(P/\bar{\rho} + E)
\end{aligned} \qquad (10)$$

$$P = (\gamma - 1)\left[\bar{\rho}\,E - \tfrac{1}{2}\,\bar{\rho}\,(U^2 + V^2)\right]$$
$$p' = (\gamma - 1)\left[\tilde{e} - (\tilde{u}\,U + \tilde{v}\,V) + \tfrac{1}{2}\,\rho'\,(U^2 + V^2)\right] \qquad (11)$$

In these expressions, $e$ is the total energy per unit mass.

## 4 Perturbation Profile and Simulation Method

We refer to the numerically applied-harmonic signal in the density, axial velocity, and pressure at the inflow near the nozzle exit by perturbations. This is in contrast the developing waves of the density, axial and radial velocities, and pressure that propagate in the airflow, which we refer to them by fluctuations. Whereas the perturbations are prescribed, the fluctuations are obtained from solving the governing system described in the previous section. We introduce a small hydrodynamic (non-propagating) perturbation wave at the inflow of the computational domain. It is specified from the centerline to a distance of two nozzle diameters. This perturbations has the following wave form

$$[\rho'\ u'\ p']^T = [\hat{\rho}(r)\ \hat{u}(r)\ \hat{p}(r)]^T \cos(\omega t) \qquad (12)$$

where $\hat{\rho}(r)$, $\hat{u}(r)$, and $\hat{p}(r)$ are the amplitudes of the perturbation wave. They have Gaussian profiles, whose nondimensional forms are shown in Fig.5. We use $R_j$ =5 mm, $\bar{\rho}_j$ =0.1114 kg/m$^3$, $U_j$ =525 m/s, and $\bar{\rho}_j U_j^2$ =30,700 N/m$^2$ for nondimensionalization throughout this study. The Gaussian profile has its maximum at $r = 0.7836\,R_j$. This is in agreement with the observed feature in such jets that the maximum intensity of the travelling waves in the axial direction is near but not at the centerline. Ali et al. [26] used similar Gaussian profile as inflow perturbation for a subsonic jet with a peak location at $r = R_j$. The choice of the current profile is determined from requiring its numerical radial derivative, using the finite difference scheme we use and discuss later, to vanish at the centerline. The perturbation amplitudes are small enough that the linear assumption implied in the governing equations is plausible. They are related by $\hat{\rho} = \hat{p}/C_G^2$ and $\hat{u} = \hat{p}/(\bar{\rho}_G C_G)$, where $\bar{\rho}_G$ and $C_G$ are the density and speed of sound in the mean flow at the peak of the Gaussian profile, respectively. The nondimensional angular frequency $\omega$ is the parameter of the perturbation in this study and will be varied to examine its effect on the fluctuations and the acoustic waves.

We use the extended MacCormack scheme of Gottlib and Turkel [27], which is $O(\Delta t^2, \Delta x^4)$. It is an extension of the classical $O(\Delta t^2, \Delta x^2)$ MacCormack predictor-corrector scheme, which was first introduced in 1969 and applied to solve the Navier-Stokes equations governing a compressible, non-heat-conducting viscous fluid in cylindrical coordinates. We choose this scheme for its relative low-computational cost in spite of its accuracy, and the continuation of using it in numerical studies of acoustics problems [28-31], thereby attesting its favorable performance. Sankar et al. [32] compared this scheme to the third order upwind scheme [33], which is $O(\Delta t^2, \Delta x^3)$ and the Lele's operator compact scheme [34], which is $O(\Delta t^2, \Delta x^4)$ and considered two aeroacoustics-related problems: acoustics field of a pulsating monopole and the diffraction and scattering of a traveling plane wave by a circular cylinder. Both problems have analytical solutions. They solved the 2D Cartesian linearized Euler equations over polar grids. The metrics of transformation were computed numerically. Radial stretching was used in the second problem. They found that the upwind scheme is the most complex and computationally expensive scheme, requiring twice the CPU time of the other schemes. The extended MacCormack scheme offered high formal spatial accuracy. They expected it to play a useful role in computational-aeroacoustics research. Snyder and Scott [35] also studied two aeroacoustics-related benchmark problems, but they are nonlinear and using the 1D Euler equations. The problems are a Gaussian pulse and a shock tube, which have approximate analytical solutions. They compared the performance of 6 different schemes, which are: MacCormack scheme $O(\Delta t^2, \Delta x^2)$, Lerat-Peyret Generalized scheme $O(\Delta t^2, \Delta x^2)$ [36], extended MacCormack scheme $O(\Delta t^2, \Delta x^4)$, Bayliss scheme $O(\Delta t^2, \Delta x^6)$ [37], Tam and Webb scheme $O(\Delta t^3, \Delta x^4)$ [38], and a new overset scheme $O(\Delta t^2, \Delta x^2)$ in which a grid zone moves to track a local flow feature like a shock. In terms of the CPU time per iteration per grid point, the MacCormack scheme was at the top of the list (fastest) and the Tam and Webb scheme was at the bottom (slowest). The ratio of the required CPU time between them was 2.55. The extended MacCormack was the fourth with 1.36 times the CPU time of the MacCormack scheme. The Lerat-Peyret and the new overset schemes were the second and third with 1.15 and 1.18 times the CPU time of the MacCormack scheme, respectively. The Bayliss scheme was the fifth with 1.62 times the CPU time of the MacCormack scheme. Therefore, lower-order schemes can outperform higher-order ones in terms of the required computational time for a specified level of accuracy. The extended MacCormack scheme outperformed the Tam and Webb scheme and was competitive with the Bayliss scheme, but the latter exhibited spurious oscillations behind the shock.

In the extended MacCormack scheme, the differential operator in the governing equations is split into four one-dimensional operators: two radial and two axial. They are applied in a symmetric way to avoid biasing of the solution, with the following sequence: first axial operator (backward difference for the predictor step followed by forward difference for the corrector step), first radial operator (backward difference for the predictor step followed by forward difference for the corrector step), second radial operator (forward difference for the predictor step followed by backward difference for the corrector step), and second axial operator (forward difference for the predictor step followed by backward difference for the corrector step). Integrating these four operators advances the solution by two time steps. Details about the scheme can be found in the studies of Ali et al. [26], Mankbadi et al. [39] and Marzouk [40]. The last reference includes validations of the method of the simulations we present here.

The boundary conditions are very important in acoustic prediction. These conditions should avoid nonphysical oscillations [41], which can deteriorate the computed solution. The boundary conditions should allow the flow and acoustic waves to leave the computational domain with as small reflection as possible. The outflow treatment here is based on the asymptotic analysis of the linearized Euler equations as given by Tam and Webb [38] and the pressure condition of Hariharan and Hagstrom [42]. The inflow treatment is based on the method of characteristics [43]. Acoustic radiation [38] and axisymmetry treatments are used for the rest of the boundary. The boundary treatments and their zones are illustrated in Fig.6. In the following, we illustrate the inflow treatment in some detail because it also accounts for the perturbation.

The full or linearized Euler equations support three types of waves: acoustic, vorticity, and entropy. The acoustic waves propagate with a velocity equal to the vector sum of the forward and backward speed of sound and the mean flow velocity. The vorticity and entropy waves are convected downstream at the same speed and direction of the mean flow. Acoustic waves are isotropic, whereas entropy and vorticity waves are highly directional. In 2D, there are four characteristic variables whose local time derivatives form the temporal growth of these waves. The main idea is to find expressions for these local time derivatives in terms of spatial derivatives of the resolved variables so that these characteristic variables can be computed, and after some manipulations, the local time derivatives of the unknown variables at the boundary are evaluated. Then, the contribution of the applied perturbation is added and the overall derivatives are integrated numerically in time. The boundary treatment applies to the primitive variables, which requires transforming the equations from the conservative form into the primitive one, and also transforming the obtained local time derivatives of the primitive variables into the conservative counterparts.

For an axial sweep of Eq.(7), one would be interested in the following 1D equation

$$\frac{\partial q'}{\partial t} + \frac{\partial F'}{\partial x} = 0 \quad (13)$$

Introducing the fluctuating primitive variables: $w' = \begin{bmatrix} \rho' & u' & v' & p' \end{bmatrix}^T$ and the Jacobians $Jq \equiv \partial q'/\partial w'$ and $JF \equiv \partial F'/\partial w'$, Eq.(13) becomes

$$\frac{\partial w'}{\partial t} + A \frac{\partial w'}{\partial x} = 0 \quad (14)$$

where $A \equiv Jq^{-1} JF = \begin{bmatrix} U & \bar{\rho} & 0 & 0 \\ 0 & U & 0 & 1/\bar{\rho} \\ 0 & 0 & U & 0 \\ 0 & \bar{\rho}C^2 & 0 & U \end{bmatrix}$

and $C = \sqrt{\gamma R_{air} T}$ is the speed of sound in the mean flow. With the similarity transformation $A = S \Lambda S^{-1}$, Eq.(14) becomes

$$\frac{\partial w'}{\partial t} + S \Lambda S^{-1} \frac{\partial w'}{\partial x} = 0 \quad (15)$$

where $\Lambda$ is a diagonal matrix containing the eigenvalues of $A$: $\lambda_i$, and the columns of $S$ are the right eigenvectors of $A$ whereas the rows of $S^{-1}$ are the transposed left eigenvectors of $A$: $l_i^T$, where $l_i^T A = \lambda_i l_i^T$. So, we have

$$\Lambda \equiv \begin{bmatrix} \lambda_1 & 0 & 0 & 0 \\ 0 & \lambda_2 & 0 & 0 \\ 0 & 0 & \lambda_3 & 0 \\ 0 & 0 & 0 & \lambda_4 \end{bmatrix} = \begin{bmatrix} U-C & 0 & 0 & 0 \\ 0 & U & 0 & 0 \\ 0 & 0 & U & 0 \\ 0 & 0 & 0 & U+C \end{bmatrix} \quad (16)$$

$$\begin{aligned} l_1^T &= \begin{bmatrix} 0 & -\bar{\rho}C & 0 & 1 \end{bmatrix} \\ l_2^T &= \begin{bmatrix} -C^2 & 0 & 0 & 1 \end{bmatrix} \\ l_3^T &= \begin{bmatrix} 0 & 0 & 1 & 0 \end{bmatrix} \\ l_4^T &= \begin{bmatrix} 0 & \bar{\rho}C & 0 & 1 \end{bmatrix} \end{aligned} \quad (17)$$

The eigenvalues $\lambda_1$ and $\lambda_4$ are in fact the propagation speeds of the forward and backward acoustic waves, respectively, and the equal eigenvalues $\lambda_2$ and $\lambda_3$ are the propagation velocities of the entropy and vorticity waves, respectively. Pre-multiplying Eq.(15) by $S^{-1}$ gives the characteristic form of Eq.(13), whose rows are four 1D characteristic equations, namely

$$S^{-1} \frac{\partial w'}{\partial t} = -\Lambda S^{-1} \frac{\partial w'}{\partial x} \quad (18)$$

This also defines a vector of local time derivatives of the characteristic variables as

$$\begin{aligned} R &\equiv S^{-1} \frac{\partial w'}{\partial t} \\ R_i &\equiv l_i^T \frac{\partial w'}{\partial t}; \quad i = 1, 2, 3, 4 \end{aligned} \quad (19)$$

So that

$$R_1 \equiv \frac{\partial p'}{\partial t} - \bar{\rho} C \frac{\partial u'}{\partial t}$$
$$R_2 \equiv \frac{\partial p'}{\partial t} - C^2 \frac{\partial \rho'}{\partial t}$$
$$R_3 \equiv \frac{\partial v'}{\partial t} \quad (20)$$
$$R_4 \equiv \frac{\partial p'}{\partial t} + \bar{\rho} C \frac{\partial u'}{\partial t}$$

Physically, $R_1$ and $R_2$ are the local time derivatives of the characteristic variables $\chi_1 \equiv p' - \bar{\rho} C u'$ and $\chi_4 \equiv p' + \bar{\rho} C u'$, which correspond to the backward and forward acoustic waves, respectively. Similarly, $R_2$ and $R_3$ are the local time derivatives of the characteristic variables $\chi_2 \equiv p' - C^2 u'$ and $\chi_3 \equiv v'$, which correspond to the entropy and vorticity waves, respectively. To find numerical values of these local time derivatives, we use the characteristic form in Eq.(18) and the definition in Eq.(19) to express these quantities in terms of spatial (axial) derivatives as

$$R_1 = -(U-C)\left(\frac{\partial p'}{\partial x} - \bar{\rho} C \frac{\partial u'}{\partial x}\right)$$
$$R_2 = -U\left(\frac{\partial p'}{\partial x} - C^2 \frac{\partial \rho'}{\partial x}\right)$$
$$R_3 = -U \frac{\partial v'}{\partial x} \quad (21)$$
$$R_4 = -(U+C)\left(\frac{\partial p'}{\partial x} + \bar{\rho} C \frac{\partial u'}{\partial x}\right)$$

To avoid reflection, outgoing waves are set to zero. For subsonic inflow, three characteristics $(\chi_2, \chi_3, \text{and } \chi_4)$ are incoming whereas $\chi_1$ is outgoing. Therefore, we need three boundary conditions to be specified, namely $R_2 = 0$, $R_3 = 0$, and $R_4 = 0$. Forward difference is used to evaluate $R_1$ from interior solution from the first relation in Eq.(21). For supersonic inflow, all four characteristics are incoming. Therefore, we need four boundary conditions, which are $R_i = 0$ with $i = 1, 2, 3,$ and $4$.

After obtaining values (zero or nonzero) for $R_i$, they are used to find the local time derivatives of the fluctuating primitive variables at the inflow boundary using Eq.(20) as

$$\left.\frac{\partial \rho'}{\partial t}\right|_\chi = \frac{1}{C^2}\left(\frac{1}{2}(R_1 + R_4) - R_2\right)$$
$$\left.\frac{\partial u'}{\partial t}\right|_\chi = \frac{1}{2\bar{\rho}C}(R_4 - R_1)$$
$$\left.\frac{\partial v'}{\partial t}\right|_\chi = R_3 \quad (22)$$
$$\left.\frac{\partial p'}{\partial t}\right|_\chi = \frac{1}{2}(R_1 + R_4)$$

Up to this point, we did not account for the perturbations in Eq.(12). However, with the aforementioned boundary treatment, the inclusion of this perturbation is simply implemented by adding the evaluated time derivatives of the perturbation signal

$$\left[\left.\frac{\partial \rho'}{\partial t}\right|_\omega \quad \left.\frac{\partial u'}{\partial t}\right|_\omega \quad \left.\frac{\partial p'}{\partial t}\right|_\omega\right]^T = \\ -\omega[\hat{\rho}(r) \quad \hat{u}(r) \quad \hat{p}(r)]^T \sin(\omega t) \quad (23)$$

to those obtained from the method of characteristics in Eq.(22). It should be mentioned here that with the operator splitting in Eq.(13), the inflow and outflow treatments are implemented only during the axial sweeps (i.e., when integrating the axial differential operators). Similarly, the axisymmetry treatment is implemented only during the radial sweeps (i.e., when integrating the radial differential operators).

We use an orthogonal grid with equal axial spacing of $0.167 R_j$ and nonuniform radial spacing varying from $0.02 R_j$ to $0.167 R_j$ as illustrated in Fig.7, which shows the distribution of the radial spacing versus the grid index. The smallest radial spacing takes place from $r = 0.8 R_j$ to $r = 1.2 R_j$, where steep radial variations are expected. This grid density is adequate and allows good resolution for wavelengths smaller than the nozzle diameter. The CFL number is set to 0.3, resulting in a nondimensional time step of 8.6e-3.

## 5 Results

We carried out the simulation for the problem described in the preceding sections for different perturbation frequencies: $\omega = \pi/20$, $4\pi/20$, and $6\pi/20$. They correspond to Strouhal numbers of 0.1, 0.4, and 0.6, respectively. These values span a range of special interest at which the plume (potential core and initial mixing layer) is very responsive to instability waves [17, 19, 44]. The simulation yields the 2D fluctuation fields which are dominated by coherent waves synchronized with the perturbation. Because the jet is ideally-expanded, no shock cells develop, which serves the aim of this study by focusing on the organized fluctuation waves, particularly the acoustic ones, due to turbulent mixing in the airflow, and how they respond to the applied perturbation.

In Fig.8, a typical axial profile of the root mean square of the fluctuating pressure $p'_{rms}$ is shown at $x = 20R_j$. The inverse proportionality with $r$ indicates the ability of the simulation to resolve the acoustic field. In Fig.9, the instantaneous 2D field of $p'$, after steady state is attained, for the different perturbation frequencies are compared. These results show that increasing the perturbation frequency causes the noise generation region to move upstream. For the low frequency, the propagation of the pressure waves is dominant in the axial direction. As the frequency increases, the oblique propagation is intensified. There is no reflection at the boundaries, which demonstrates the excellent performance of the adopted simulation method. The sound directivity curves, sound pressure levels (SPL) versus the observer angle: $\arctan(r/x)$, at different values of the polar coordinate: $\sqrt{x^2 + r^2}$ are compared in Fig.10. The differences at the large values of the polar coordinate ($42R_j$ and $48R_j$) are different from those at smaller values ($30R_j$ and $36R_j$). For all frequencies, the SPL increases and then decreases with the observer angle at the larger values of the polar coordinate. However, as the frequency increases, the peak SPL is shifted toward higher observer angles. This is a consequence of the changes in the directional pattern of the pressure waves. For low perturbation frequency, the SPL reaches high values (near 160 dB) at small observer angles and small values of the polar coordinate, and the SPL decreases monotonically as the angle increases. When the frequency increases, the SPL behavior at these smaller values of the polar coordinate becomes similar to the one at the larger values, exhibiting a peak corresponding to the oblique propagation. In terms of the strength of the pressure waves, we consider the case of a polar coordinate=$48R_j$ for the purpose of comparing the effect of the perturbation. The peak for this curve corresponds to SPL=145 dB for the low frequency, which is 4dB smaller (corresponding to attenuation by a factor of 1.6) than the peak for the other frequencies. Therefore, the perturbation frequency not only change the propagating pressure waves qualitatively, but also quantitatively.

In Fig.11, we present the axial profiles of nondimensional $p'$ at different values of $r$ varying from $R_j$ to $30R_j$ for each perturbation case. Increasing the frequency causes $x$ value at which the amplitude of the pressure wave reaches its maximum to decrease (from $46R_j$ for the low frequency to $11R_j$ for the high frequency). There is also an influence of the perturbation on the radial decay in the pressure waves. This is illustrated by comparing the waves at radial distances of $R_j$ and $5R_j$ for the different perturbation cases. At the low frequency, the amplitudes at these values of $r$ are very close, indicating very weak damping. In contrast, there is strong damping in the same region for the other frequencies, where the maximum amplitude of pressure wave decreases by a factor of two.

In Figs.12-14, we compare the $p'$ in the time domain for the different perturbations at four values of $x$: $10R_j$, $20R_j$, $30R_j$, and $50R_j$, and three values of $r$: $R_j$, $5R_j$, and $10R_j$. These figures support the earlier discussion on the changes in the radial decay due to the perturbation frequency. In addition, these figures demonstrate the synchronization of the waves with the perturbation. The time periods (nondimensional) of the waves (40, 10, and 6.67 time units) are equal to the period of the corresponding perturbation. To better contrast the intensities of the pressure wave at these 12 points distributed in the domain, we use same axis limits for $p'$. Regardless of the frequency, the highest intensity is at $r = R_j$, although the intensity does not decrease significantly up to decay $r = 5R_j$ for the low-frequency case as shown in Fig.12. Considering all perturbation cases, the maximum intensity occurs at $x=10R_j$ for the high-frequency case, where the amplitude of the wave reaches a nondimensional value of 0.08. The counterpart amplitude for $\omega = 4\pi/20$ is very close. For the low frequency, the maximum amplitude occurs at $x=30R_j$ with a nondimensional value of 0.06.

In Figs.15 and 16, we examine the instantaneous 2D fields of $u'$ and $v'$, respectively, in state-state conditions for the different frequencies. There is qualitative similarity between the two fields, as suggested by linear-stability analysis. These fields are intensified only within a small region near the centreline, and the vorticity levels are very small elsewhere, indicating that the pressure fluctuations in the main part of the domain (as in the decaying curve in Fig.8) are acoustic waves. As was the case with the pressure fields, as the frequency increases, the axial extension of the fluctuations is shortened for the velocity fields, and the oblique propagation is strengthened.

## 6 Conclusion

We studied the coherent fluctuations and associated acoustic waves generated by perturbing a supersonic ideally-expanded air jet. The frequency was the parameter of the perturbation signal and we examined its influence, as it varies over a range of high wave amplification, on the fluctuations fields, with emphasis on the pressure waves. We studied the changes in the directivity pattern, the axial profiles of the pressure waves, their behavior in the time domain at different locations in the domain, and the steady-state 2D propagation of the pressure and velocity fluctuations. The mean flow is described by a set of relations based on the conservation of mass, momentum, and energy, combined with Gaussian-profile approximation for the axial velocity in the mixing layer. We numerically integrated the linearized Euler equations using the finite difference method and a set of non-reflecting boundary conditions. The input perturbation is prescribed for the density, axial velocity, and pressure at the inflow through a treatment based on the method of characteristics.

The pressure waves are synchronized with the perturbation. The applied frequency can alter the pressure waves and thus the sound field. For low frequencies, the waves propagate dominantly in the parallel direction, whereas the oblique propagation is intensive at high frequencies. The radial decay of the waves is weak for the low frequencies and the wave amplitude is small compared to the high frequencies. The sound generation region moves upstream as the perturbation frequency is increased, and the axial extension of the propagating fluctuations is noticeably shortened. The steady-state fields of the velocity waves are similar to those for the pressure one. The results show that the applied perturbation frequency can be viewed as a device to control the sound levels by changing the pressure propagation pattern, thereby keeping certain regions, where noise needs to be minimized, away from the strong wavefronts.

*References:*
[1]  A. T. Jones, Singing Flames, *The Journal of the Acoustical Society of America*, Vol.16, No.4, 1945, pp. 254-266.
[2]  A. Thomas and G. T. Williams, Flame Noise: Sound Emission from Spark-Ignited Bubbles of Combustible Gas, *Proceedings of the Royal Society of London. Series A: Mathematical and Physical Sciences*, Vol.294, No.1439, 1966, pp. 449-466.
[3]  A. W. Morley, Estimation of Aeroplane Noise Level: Some Empirical Laws with an Account of the Present Experiments on Which They Are Based, *Aircraft Engineering and Aerospace Technology*, Vol.11, No.123, 1939, pp. 187-189.
[4]  M. J. Lighthill, On Sound Generated Aerodynamically: I- General Theory, *Proceedings of the Royal Society of London. Series A: Mathematical and Physical Sciences*, Vol.211, No.1107, 1952, pp. 564-587.
[5]  F. Owis and P. Balakumar, Jet Noise Computations Using Explicit MacCormack Schemes and Nonrflecting Boundary Conditions, *AIAA Journal*, Vol.39, No.10, 2001, pp. 2019-2021.
[6]  P. Moore, H. Slot, and B. J. Boersma, Simulation and Measurement of Flow Generated Noise, *Journal of Computational Physics*, Vol.224, No.1, 2007, pp. 449-463.
[7]  C. K. W. Tam, LES for Aeroacoustics, *Proceedings of the 29th AIAA Fluid Dynamics Conference*, Albuquerque, New Mexico, June 15–18, 1998, AIAA-1998-2805.
[8]  N. Andersson and L.-E. Eriksson, Prediction of Flowfield and Acoustic Signature of a Coaxial Jet Using RANS-Based Methods and Large-Eddy Simulation, *International Journal of Aeroacoustics*, Vol.7, No.1, 2008, pp. 23-40.
[9]  D. J. Bodony and S. K. Lele, Review of the Current Status of Jet Noise Predictions Using Large-Eddy Simulation, *Proceedings of the Aerospace Science Meeting and Exhibit*, Reno, Nevada, January 9-12, 2006, AIAA-2006-0486.
[10] F. Mathey, Aerodynamic Noise Simulation of the Flow past an Airfoil Trailing-Edge Using a Hybrid Zonal RANS-LES, *Computers & Fluids*, Vol.37, No.7, 2008, pp. 836-843.

[11] E. Gröschel, W. Schröder, P. Renze, M. Meinke, and P. Comte, Noise Prediction for a Turbulent Jet Using Different Hybrid Methods, *Computers & Fluids*, Vol.37, No.4, 2008, pp. 414-426.

[12] F. Q. Hu, On Absorbing Boundary Conditions for Linearized Euler Equations by Perfectly Matched Layer, *Journal of Computational Physics*, Vol.129, No.1, 1996, pp. 201-219.

[13] K. Sreenivas and D. L. Whitfield, Time- and Frequency-Domain Numerical Simulation of Linearized Euler Equations, *AIAA Journal*, Vol.36, No.6, 1998, pp. 968-975.

[14] I. Ali, S. Becker, J. Utzmann, and C.–D. Munz, Aeroacoustic Study of a Forward Facing Step Using Linearized Euler Equations, *Physica D*, Vol.237, No.14-17, 2007, pp. 2184-2189.

[15] P. Freymuth, On Transition in a Separated Laminar Boundary Layer, *Journal of Fluid Mechanics*, Vol.25, No.4, 1966, pp. 683-704.

[16] D. Ru-Sue Ko, T. Kubota, and R. Lees, Finite Disturbance Effect on the Stability of a Laminar Incompressible Wake Behind a Flat Plate, *Journal of Fluid Mechanics*, Vol.40, No. 2, 1970, pp. 315-367.

[17] Y. Y. Chan, Spatial Waves in Turbulent Jets, *Physics of Fluids*, Vol.17, No.1, 1974, pp. 46-53.

[18] D. K. McLaughlin, G. L. Morrison, and T. R. Troutt, Experiments on the Instability Waves in a Supersonic Jet and Their Acoustic Radiation, *Journal of Fluid Mechanics*, Vol.69, 1975, pp. 73-95.

[19] T. R. Troutt, Measurements on the Flow and Acoustic Properties of a Moderate-Reynolds-Number Supersonic Jet, *Ph.D. Dissertation*, Oklahoma State University, Oklahoma, 1978.

[20] P. J. Morris and C. K. W. Tam, Near- and Far-Field Noise from Large-Scale Instabilities of Axisymmetric Jets, *Proceedings of the 4th AIAA Aeroacoustics Conference*, Atlanta, Georgia, October 3-5, 1977, AIAA-1977-1351.

[21] R. R. Mankbadi, S. H. Shih, R. Hixon, and L. A. Povinelli, Direct Computation of Jet Noise Produced by Large-Scale Axisymmetric Structures, *Journal of Propulsion and Power*, Vol.16, No.2, 2000, pp. 207-215.

[22] P. Lew, A. Uzun, G. A. Blaisdell, and A. S. Lyrintzis, Effects of Inflow Forcing on Jet Noise Using Large Eddy Simulation, *Proceedings of the 42nd AIAA Aerospace Science Meeting and Exhibit*, 2004, AIAA-2004-0516.

[23] C. K. W. Tam and D. E. Burton, Sound Generated by Instability Waves of Supersonic Flows. Part2. Axisymmetric Jets, *Journal of Fluid Mechanics*, Vol.138, 1984, pp. 273-295.

[24] T. R. Trout and D. K. McLaughlin, Experiments on the Flow and Acoustic Properties of a Moderate-Reynolds-Number Supersonic Jet, *Journal of Fluid Mechanics*, Vol.116, 1982, pp. 123-156.

[25] M. Idres, Computational Modeling of Airborne Noise Demonstrated via Benchmarks, Supersonic Jet, and Railway Barrier, *Ph.D. Dissertation*, Old Dominion University, Virginia, 1990.

[26] A. Ali, A. Hamed, R. Hixon, R. Mankbadi, A. Mobarek, and M. Rizk, Effect of Inflow Treatment on Acoustic Radiation from Large-Scale Structure in a Round Jet, *Proceedings of the 37th AIAA Aerospace Science Meeting and Exhibit*, Reno, Nevada, January, 11-14, 1999, AIAA-1999-0083.

[27] D. Gottlib, and E. Turkel, Dissipative Two-Four Method for Time Dependent Problems, *Mathematics of Computation*, Vol.30, No.136, 1976, pp. 703-723.

[28] Z. Zhu, Application of Finite Difference Method Adopted in Fluid Mechanics to Sound Propagation in Ducts, *Acta Mechanica Solida Sinica*, Vol.3, No.4, 1990, pp. 457-468.

[29] C. Tsingas, A. Vafidis, and E. R. Kanasewich, Elastic Wave Propagation in Transversely Isotropic Media Using Finite Differences, *Geophysical Prospecting*, Vol.38, No.8, 1990, pp. 933-949.

[30] K. Schneider, O. Roussel, and M. Farge, Coherent Vortex Simulation (CVS) of Compressible Turbulent Mixing Layers Using Adaptive Multiresolution Methods, *Proceedings of the Annual Meeting of the Division of Fluid Dynamics of the American Physical Society*, Salt Lake City, Utah, November 18-20, 2007.

[31] S. Biringen, J. E. Howard, and R.S. Reichert, Simulation of Sonic Boom Interaction with Shear Turbulence, *Mechanics Research Communications*, Vol.32, 2005, pp. 604-609.

[32] L. N. Sankar, N. N. Reddy, and N. Hiriharan, A Comparative Study of Numerical Schemes for Aeroacoustic Applications, in *Computational Aero- and Hydro-Acoustics*, ASME Fluid Engineering Division, edited by R. R. Mankbadi et al., Vol.147, 1993, pp. 35-40.

[33] L. N. Sankar, N. N. Reddy, and N. Hariharan, A Third Order Upwind Scheme for Aero-Acoustic Applications , *Proceedings of the 31st AIAA Aerospace Sciences Meeting*, Reno,

Nevada, January 11-14, 1993, AIAA-1993-0149.

[34] S. K. Lele, Direct Numerical Simulation of Compressible Free Shear Flows, *Proceedings of the 27th AIAA Aerospace Sciences Meeting*, Reno, Nevada, January 9-12, 1989, AIAA-1989-0374.

[35] R. D. Snyder and J. N. Scott, Comparison of Numerical Schemes for the Analysis of Aeroacoustics, *Proceedings of the 37th AIAA Aerospace Science Meeting and Exhibit*, Reno, Nevada, January, 11-14, 1999, AIAA-1999-0354.

[36] R. Peyret and T. D. Taylor, *Computational Methods for Fluid Flow*, Springer Series in Computational Physics, Springer-Verlag, 1983.

[37] A. Bayliss, L. Maestrello, P. Parikh, and E. Turkel, A Fourth Order Scheme for the Unsteady Compressible Navier-Stokes Equations, *NASA Report*, 1985, ICASE-85-44.

[38] C. K. W. Tam and J. C. Webb, Dispersion-Relation-Preserving Finite Difference Schemes for Computational Acoustics, *Journal Computational Physics*, Vol.107, No.2, 1993, pp. 262-283.

[39] R. R. Mankbadi, M. E. Hayder, and L. A. Povinelli, The Structure of Supersonic Jet Flow and Its Radiated Sound, *AIAA Journal*, Vol.32, No.5, 1994, pp. 897-906.

[40] O. A. Marzouk, A Two-Step Computational Aeroacoustics Method Applied to High-Speed Flows, *Noise Control Engineering Journal*, Vol.56, No.5, 2008.

[41] C. K. W. Tam and P. J. Morris, The Radiation of Sound by the Instability Waves of a Compressible Plane Turbulent Shear Layer, *Journal of Fluid Mechanics*, Vol.98, 1980, pp. 349-381.

[42] S. I. Hariharan and T. Hagstrom, Far Field Expansion for Anisotropic Wave Equations, *Proceedings of the IMACS Symposium on Computational Acoustics*, edited by D. Lee, A. Cakmak, and R. Vichnevtsky, Amsterdam, North Holland, Vol.2, 1990, pp. 283-294.

[43] K. W. Thompson, Time-Dependent Boundary Conditions for Hyperbolic Systems II, *Journal of Computational Physics*, Vol.89, 1990, pp. 439-461.

[44] S. C. Crow and F. H. Champagne, Orderly Structure in Jet Turbulence, *Journal of Fluid Mechanics*, Vol.48, 1971, pp. 547-591.

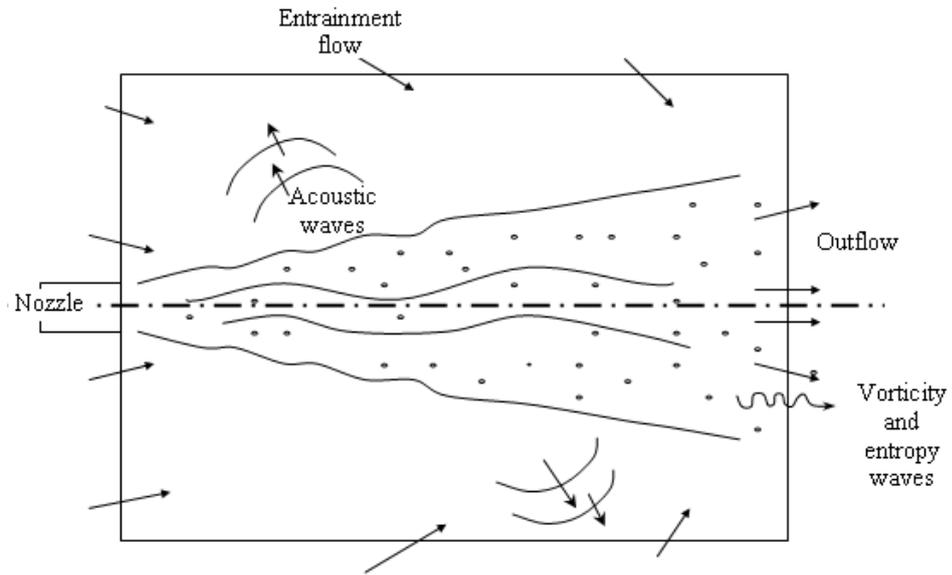

**Figure 1: Schematic diagram of a jet**

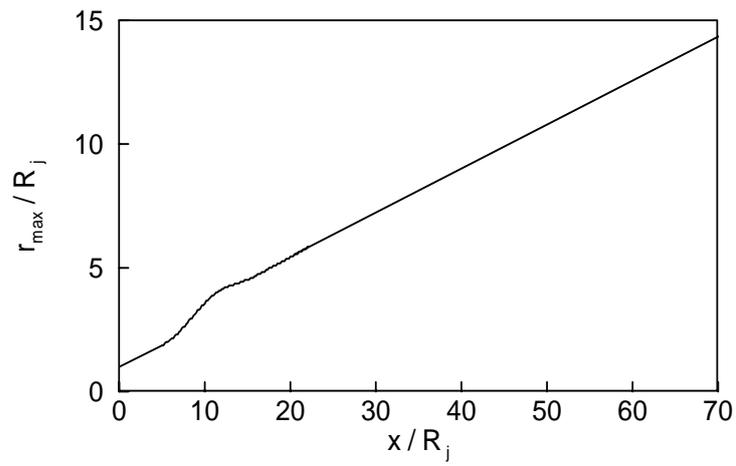

**Figure 2: Axial profile of $r_{max}(x)$**

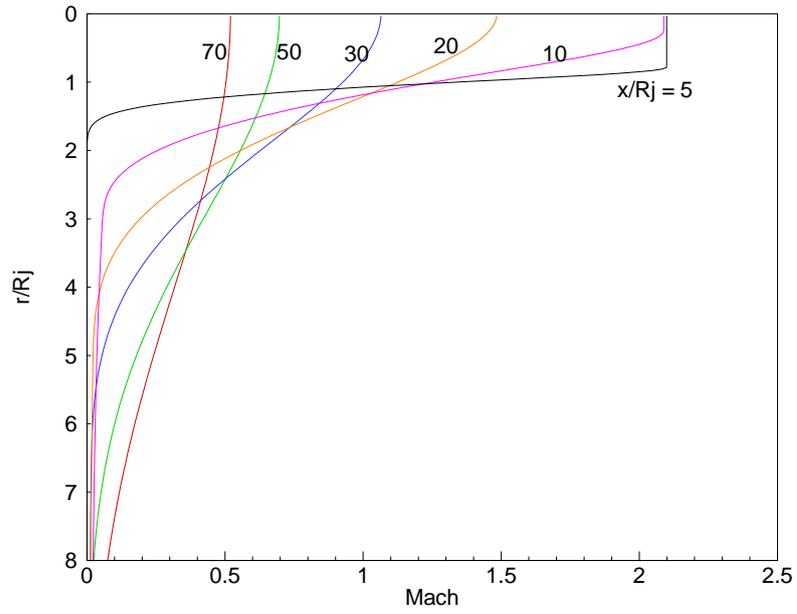

**Figure 3: Radial profiles of $M$**

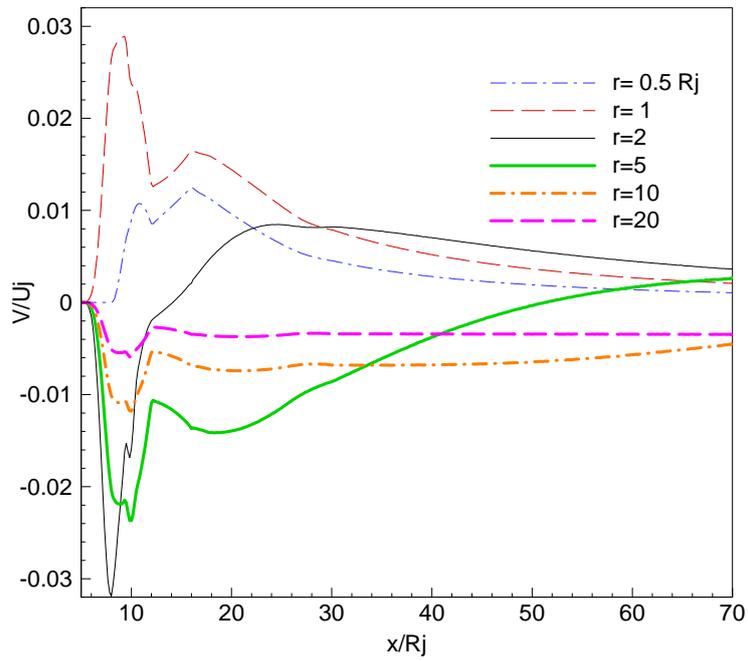

**Figure 4: Axial profiles of $V$ relative to $U_j$**

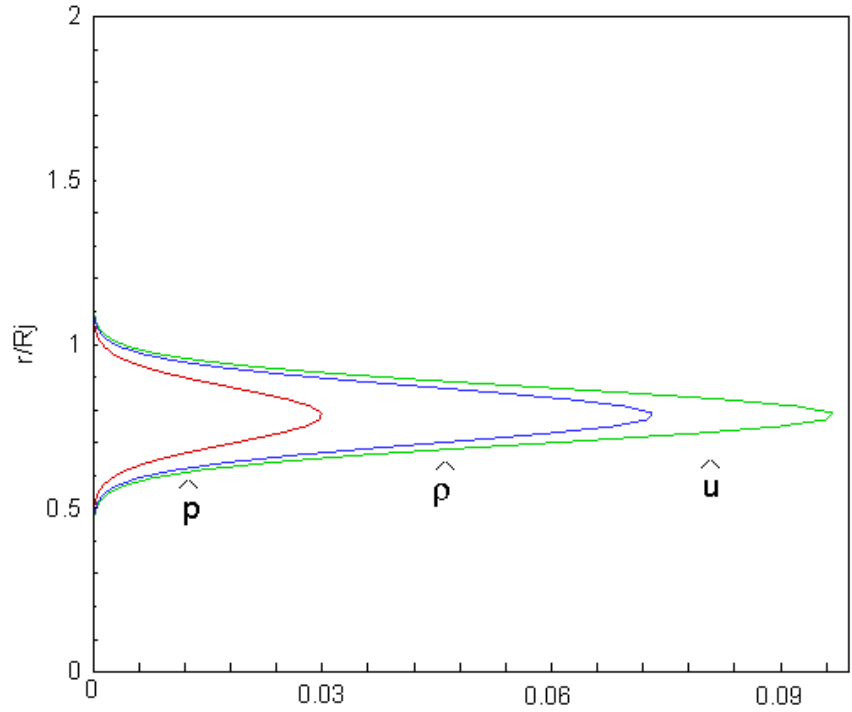

**Figure 5: Nondimensional amplitudes of the perturbation**

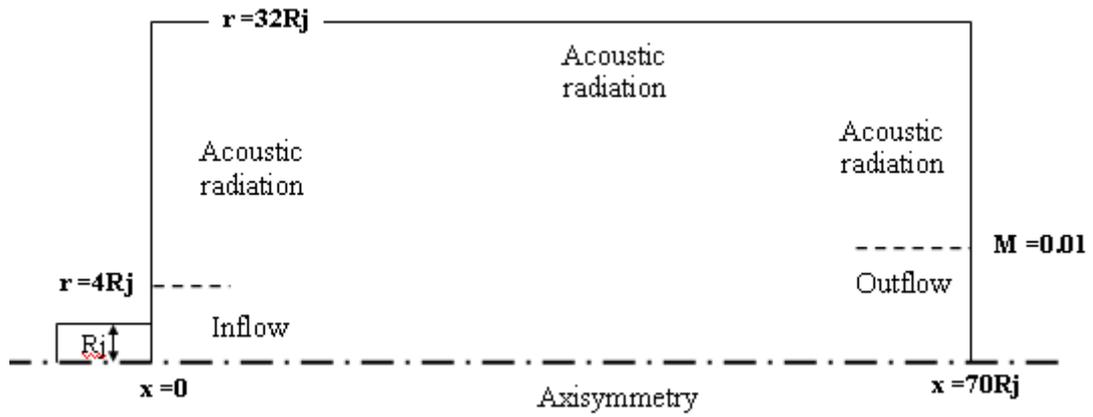

**Figure 6: Problem domain and boundary treatments**

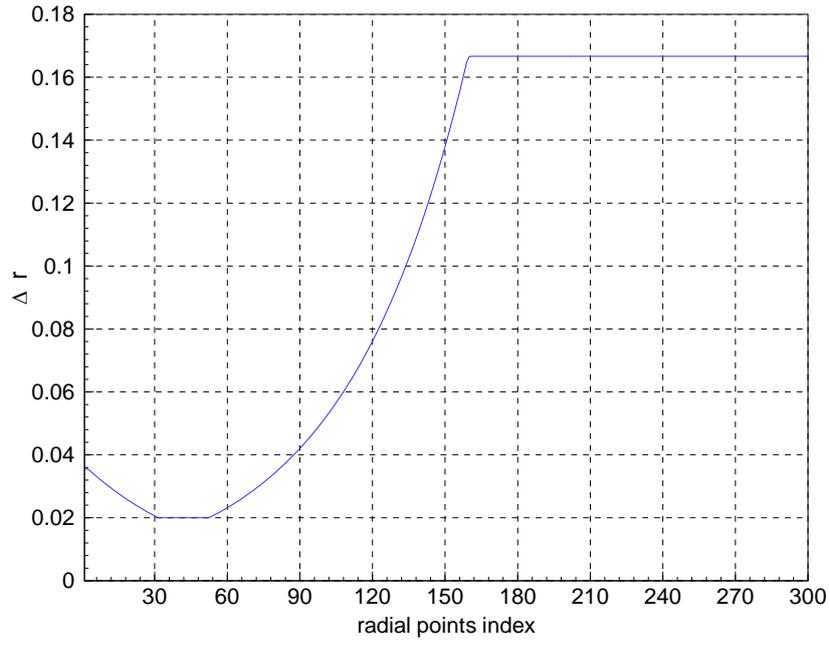

**Figure 7: Nonuniform grid spacing in the radial direction**

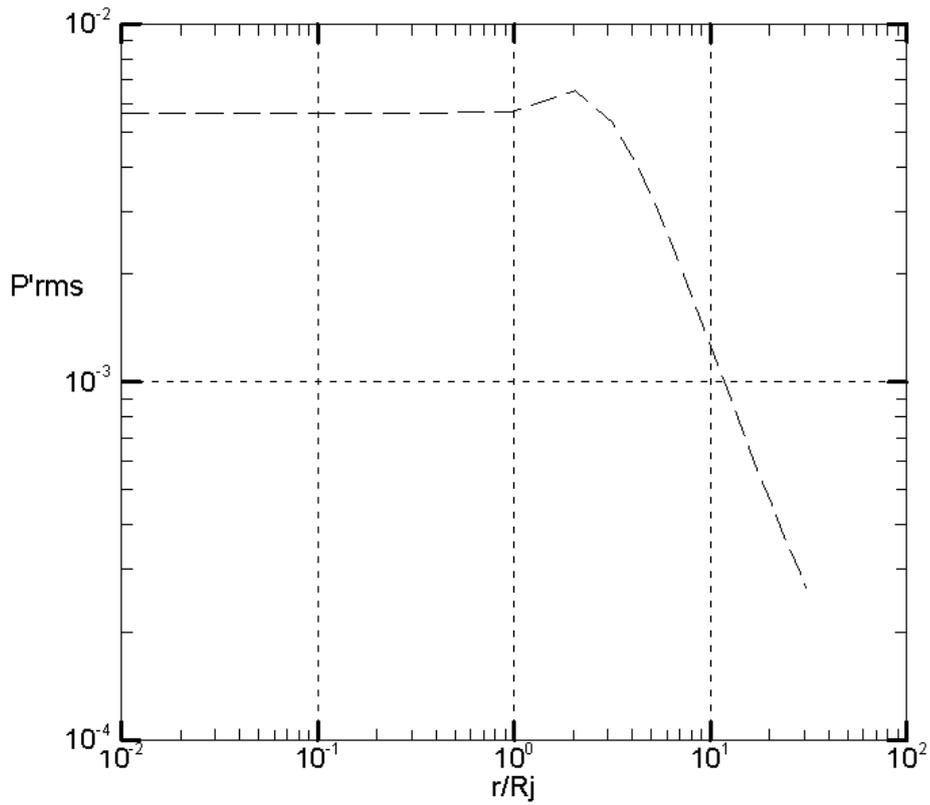

**Figure 8: Axial profile of nondimensional $p'_{rms}$ at $x = 20\,R_j$**

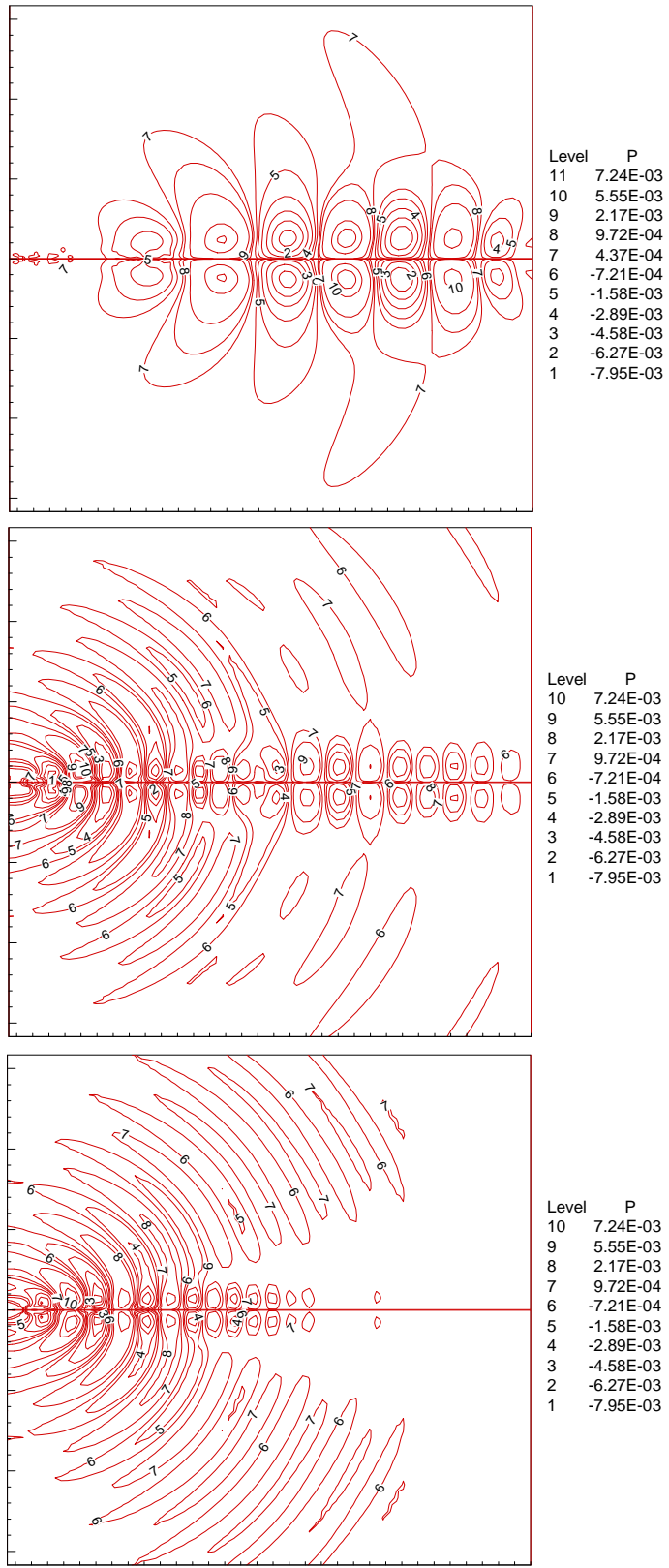

**Figure 9: Instantaneous 2D field of $p'$ for $\omega = \pi/20$ (top), $\omega = 4\pi/20$ (center), and $\omega = 6\pi/20$ (bottom)**

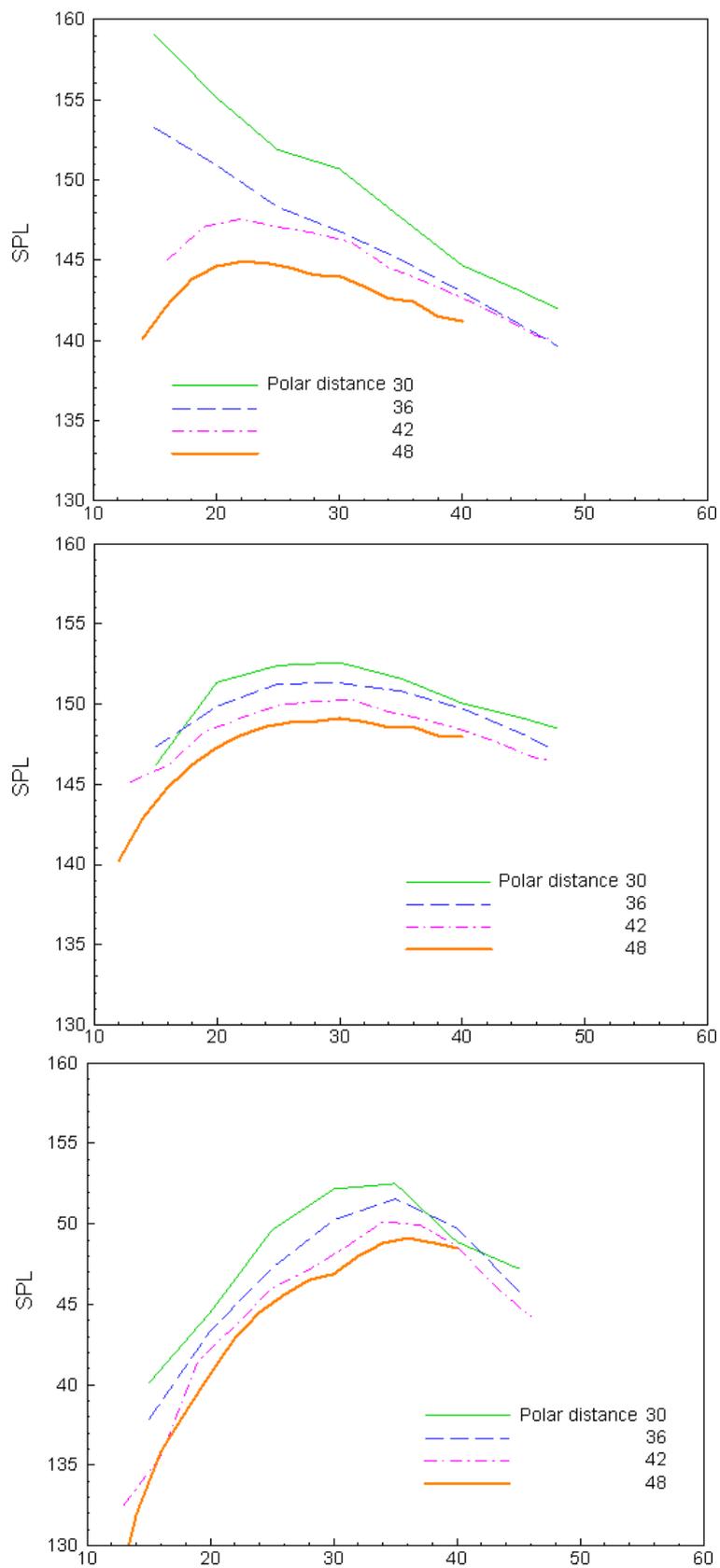

**Figure 10:** Sound directivity for $\omega = \pi/20$ (top), $\omega = 4\pi/20$ (center), and $\omega = 6\pi/20$ (bottom)

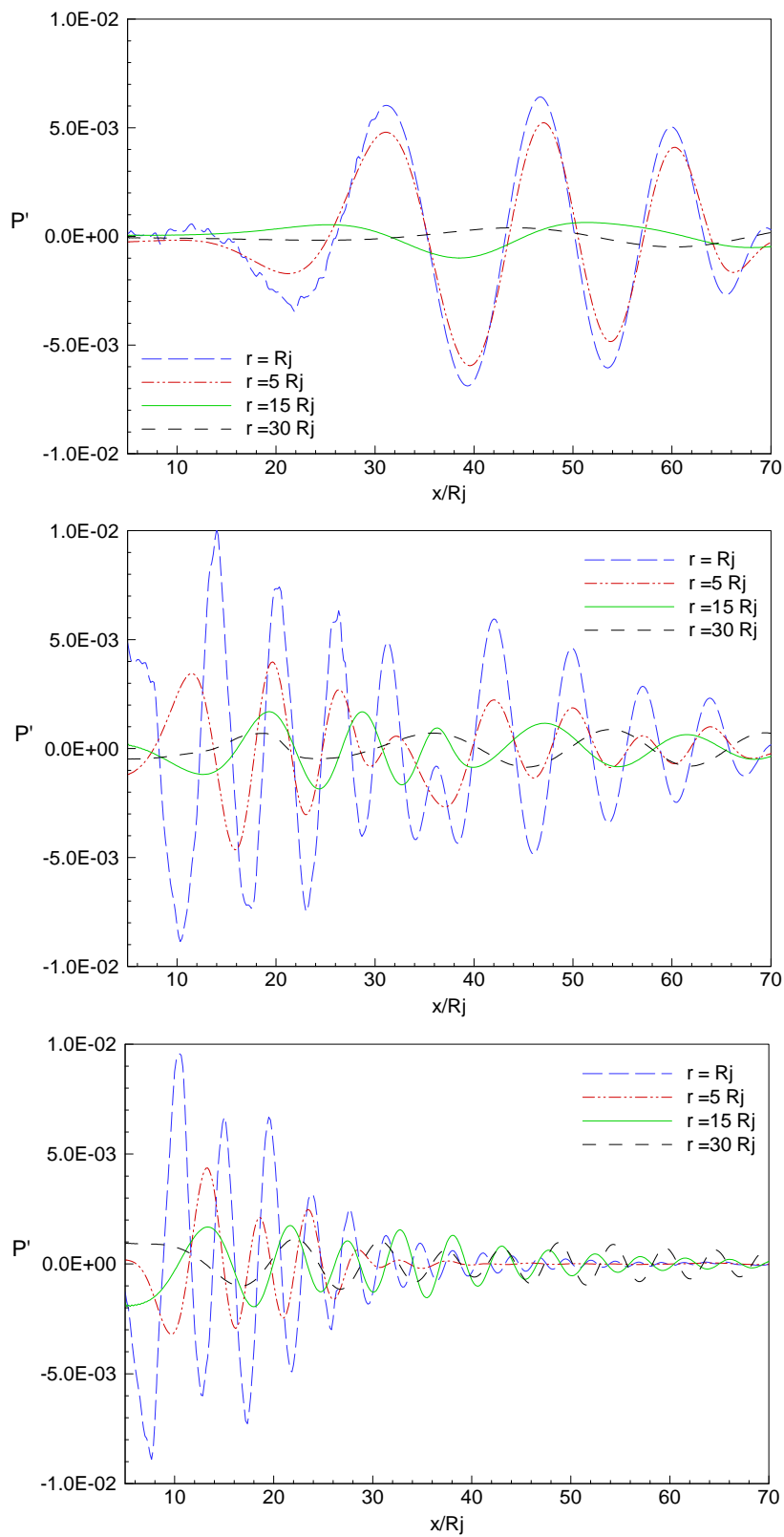

**Figure 11: Axial profiles of nondimensional $p'$ for $\omega = \pi/20$ (top), $\omega = 4\pi/20$ (center), and $\omega = 6\pi/20$ (bottom)**

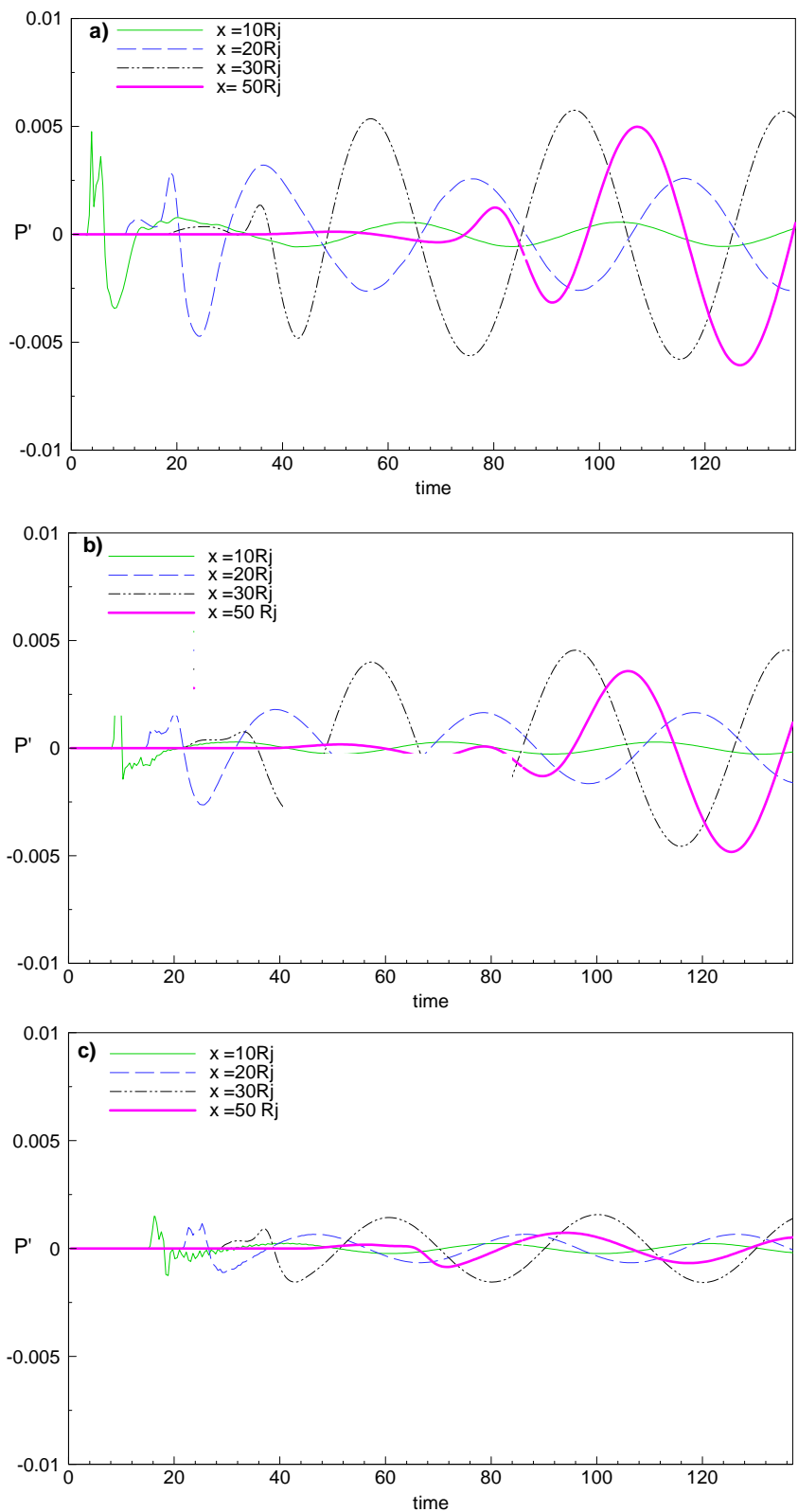

**Figure 12:** Nondimensional $p'$ versus nondimensional time for $\omega = \pi/20$ at different values of $x$ and $r = R_j$ (top), $r = 5R_j$ (center), and $r = 10R_j$ (bottom)

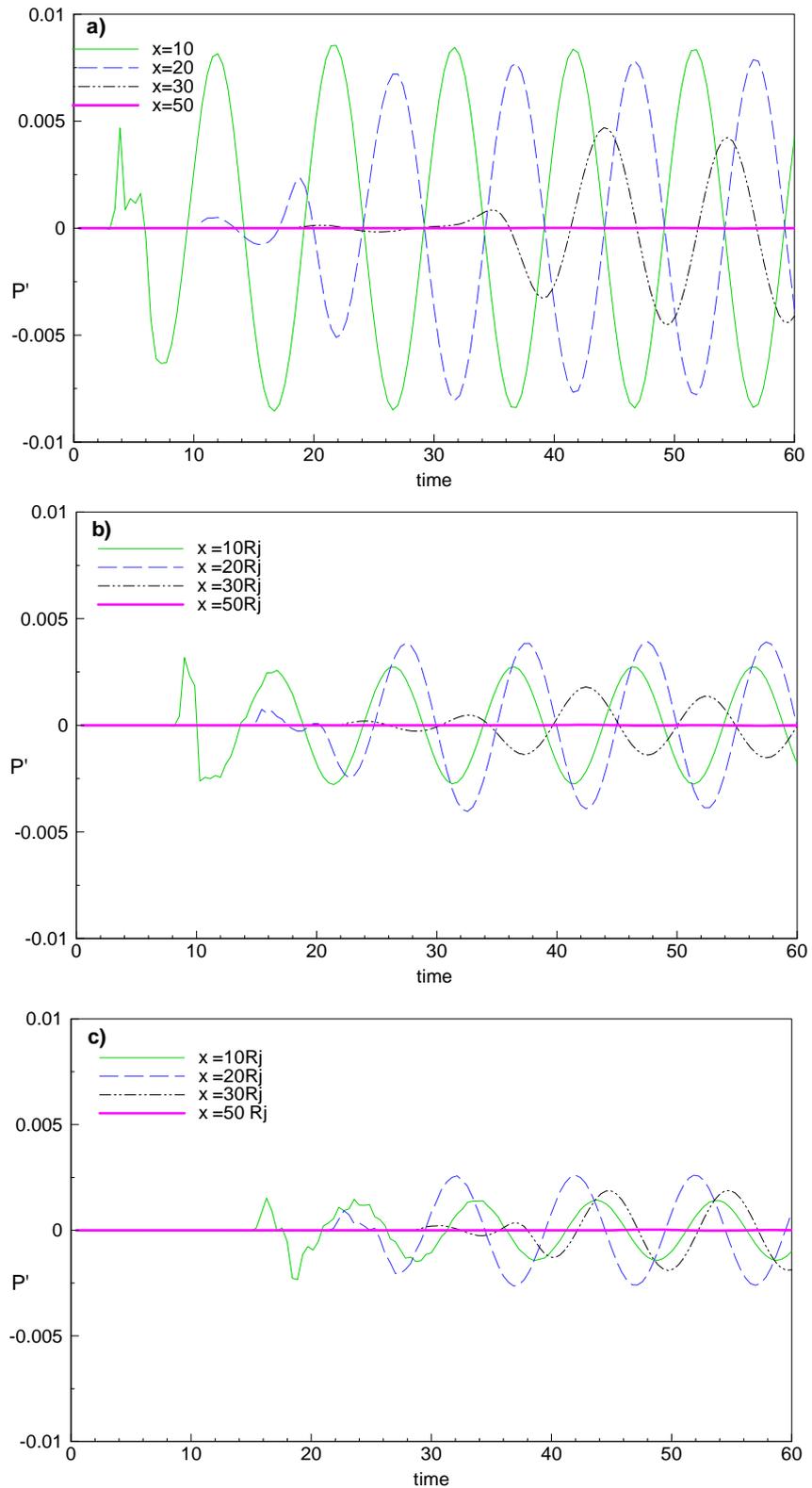

**Figure 13:** Nondimensional $p'$ versus nondimensional time for $\omega = 4\pi/20$ at different values of $x$ and $r = R_j$ (top), $r = 5R_j$ (center), and $r = 10R_j$ (bottom)

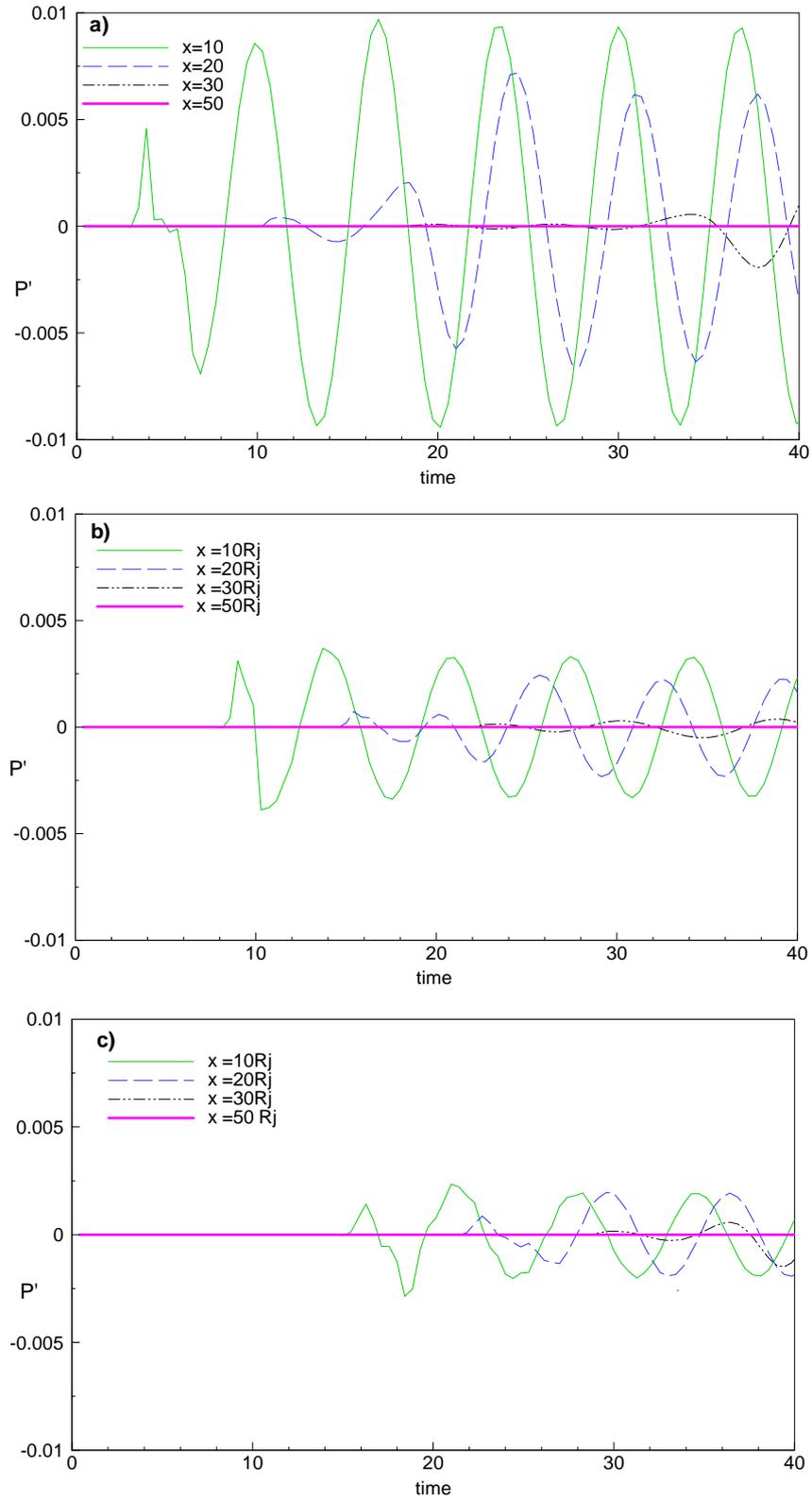

**Figure 14:** Nondimensional $p'$ versus nondimensional time for $\omega = 6\pi/20$ at different values of $x$ and $r = R_j$ (top), $r = 5R_j$ (center), and $r = 10R_j$ (bottom)

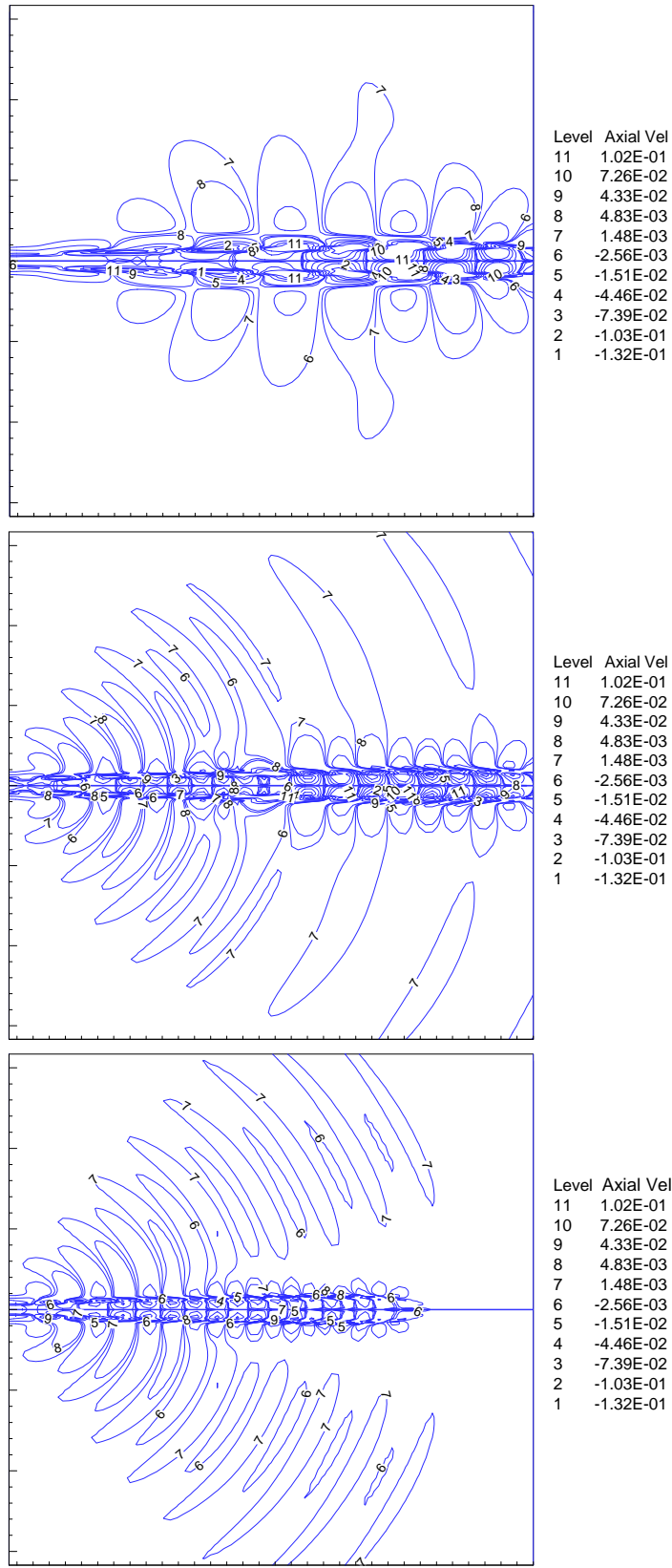

**Figure 15: Instantaneous 2D field of** $u'$ **for** $\omega = \pi/20$ **(top),** $\omega = 4\pi/20$ **(center), and** $\omega = 6\pi/20$ **(bottom)**

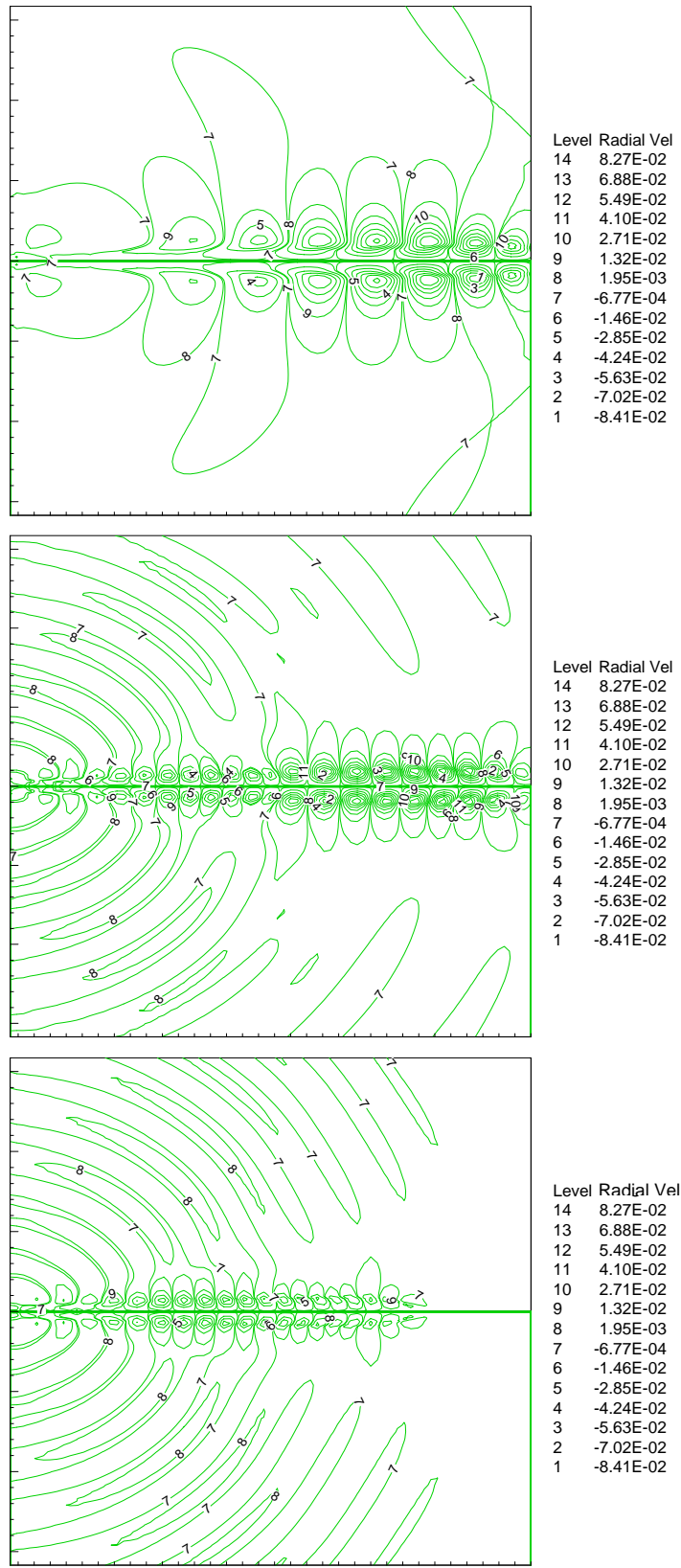

**Figure 16: Instantaneous 2D field of $v'$ for $\omega = \pi/20$ (top), $\omega = 4\pi/20$ (center), and $\omega = 6\pi/20$ (bottom)**